\documentclass[sigconf,screen,nonacm]{acmart}

\usepackage{pifont}
\usepackage{siunitx}
\usepackage{multirow}
\usepackage{subcaption}
\usepackage[linesnumbered,ruled,vlined]{algorithm2e}
\SetAlgoSkip{9pt}
\usepackage{enumitem}
\usepackage{mathtools}
\usepackage{tablefootnote}

\setcopyright{none}

\newcommand{\Finesse}{\texttt{Finesse}}

\begin{document}

\title{Finesse: An Agile Design Framework for Pairing-based Cryptography via Software/Hardware Co-Design}

\author{Tianwei Pan}
\orcid{0009-0001-6633-4076}
\authornote{Both authors contributed equally to this research.}
\affiliation{
  \institution{Beihang University}
  \city{Beijing}
  \country{China}
}
\email{pantw@buaa.edu.cn}

\author{Tianao Dai}
\orcid{0009-0000-3164-048X}
\authornotemark[1]
\affiliation{
  \institution{Beihang University}
  \city{Beijing}
  \country{China}
}
\email{daitianao@buaa.edu.cn}

\author{Jianlei Yang}
\orcid{0000-0001-8424-7040}
\authornote{Corresponding author is Jianlei Yang. Email: \url{jianlei@buaa.edu.cn}.\\
This work is supported in part by the National Natural Science Foundation of China (Grant No. 62072019), the Beijing Natural Science Foundation (Grant No. L243031), and the National Key R\&D Program of China (Grant No. 2023YFB4503704 and 2024YFB4505601).\\
The \Finesse~framework is avaiable at \url{https://github.com/BUAA-CI-LAB/Finesse}.}
\affiliation{%
  \institution{Beihang University}
  \city{Beijing}
  \country{China}
}
\email{jianlei@buaa.edu.cn}

\author{Hongbin Jing}
\orcid{0009-0004-0202-5065}
\affiliation{%
  \institution{Beihang University}
  \city{Beijing}
  \country{China}
}
\email{hongbin@buaa.edu.cn}

\author{Yang Su}
\orcid{0009-0001-3293-7709}
\affiliation{%
  \institution{Beihang University}
  \city{Beijing}
  \country{China}
}
\email{buaayangsu@buaa.edu.cn}

\author{Zeyu Hao}
\orcid{0009-0009-0175-6938}
\affiliation{%
  \institution{Beihang University}
  \city{Beijing}
  \country{China}
}
\email{withinlover@buaa.edu.cn}

\author{Xiaotao Jia}
\orcid{0000-0003-2207-6092}
\affiliation{%
  \institution{Beihang University}
  \city{Beijing}
  \country{China}
}
\email{jiaxt@buaa.edu.cn}

\author{Chunming Hu}
\orcid{0000-0003-3473-9703}
\affiliation{%
  \institution{Beihang University}
  \city{Beijing}
  \country{China}
}
\email{hucm@buaa.edu.cn}

\author{Weisheng Zhao}
\orcid{0000-0001-8088-0404}
\affiliation{%
  \institution{Beihang University}
  \city{Beijing}
  \country{China}
}
\email{weisheng.zhao@buaa.edu.cn}

\begin{abstract}
Pairing-based cryptography (PBC) is crucial in modern cryptographic applications.
With the rapid advancement of adversarial research and the growing diversity of application requirements, PBC accelerators need regular updates in algorithms, parameter configurations, and hardware design.
However, traditional design methodologies face significant challenges, including prolonged design cycles, difficulties in balancing performance and flexibility, and insufficient support for potential architectural exploration.

To address these challenges, we introduce \texttt{Finesse}, an agile design framework based on co-design methodology.
\texttt{Finesse} leverages a co-optimization cycle driven by a specialized compiler and a multi-granularity hardware simulator, enabling both optimized performance metrics and effective design space exploration.
Furthermore, \texttt{Finesse} adopts a modular design flow to significantly shorten design cycles, while its versatile abstraction ensures flexibility across various curve families and hardware architectures.

\texttt{Finesse} offers flexibility, efficiency, and rapid prototyping, comparing with previous frameworks. With compilation times reduced to minutes, \texttt{Finesse} enables faster iteration cycles and streamlined hardware-software co-design. Experiments on popular curves demonstrate its effectiveness, achieving $34\times$ improvement in throughput and $6.2\times$ increase in area efficiency compared to previous flexible frameworks, while outperforming state-of-the-art non-flexible ASIC designs with a $3\times$ gain in throughput and $3.2\times$ improvement in area efficiency.

\end{abstract}




\maketitle

\section{Introduction}

Bilinear pairing, since its formulation in modern cryptography, has been a crucial primitive for building advanced cryptographic protocols and systems. Pairings enable efficient schemes for identity-based encryption~\cite{boneh2001identity}, attribute-based encryption~\cite{sahai2005fuzzy}, short signature~\cite{boneh2004short}, SNARKs such as KZG~\cite{KZG} and Groth16~\cite{groth2016size}. As data security and privacy gain increasing attention, pairing-based cryptography plays a significant role in this context, safeguarding user data and underpinning the trust and reliability of modern digital infrastructures. Despite its benefits, pairing comes with a significant computational cost. While traditional signature schemes offer a latency of around \SI{20}{\micro\second} on desktop CPUs, pairing computations are typically 2 orders of magnitude longer~\cite{relic-toolkit}. Application needs for pairing have motivated researchers to embark on the exploration of efficient pairing acceleration techniques, targeting platforms ranging from CPUs~\cite{relic-toolkit, azar2017}, GPUs~\cite{p1363gpu}, to server-side FPGAs~\cite{sakamoto2024high} and ASIC~\cite{ASSCC2019}. Among these, FPGAs and ASICs excel primarily due to their support for domain-specific designs~\cite{hennessy2018new}, enabling customized datapaths and logic units that achieve low latency and high resource efficiency in terms of area and power for pairing accelerators.

\begin{figure}[t]
    \centering
    \includegraphics[width=\linewidth]{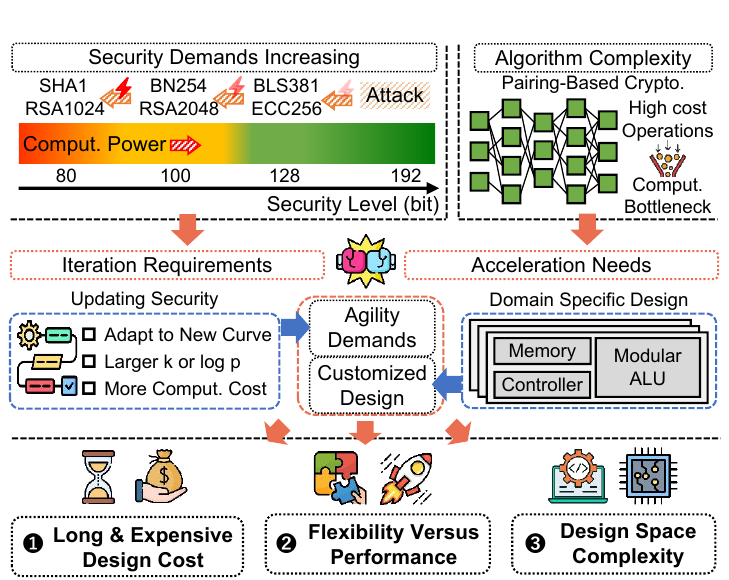}
    \caption{The challenges of PBC accelerator design.}
    \label{fig:security-needs}
\end{figure}

However, research in this area faces several critical challenges, as illustrated in \autoref{fig:security-needs}. The security level of pairings is not static; it diminishes as attack methods improve and as computational power advances~\cite{FOTIADIS201973,Barbulescu2015,Micheli2020,aranha2023survey}. The growing diversity of application requirements in pairing-based cryptography (PBC) necessitates iterative updates to maintain security and performance standards. This ongoing need for adaptation gives rise to \textbf{challenge {\ding{182}}: the PBC hardware re-engineering costs}.
 
Existing works with high performance efficiency are designed with specific pairing parameters in mind. While these designs achieve good results, they also lead to high re-engineering costs. On the other hand, works with higher flexibility suffer from low performance due to a lack of parallelism in their architecture, and their flexibility is constrained by the specific hardware implementation (e.g. difficulty supporting curves with larger bit widths), which also results in unavoidable re-engineering costs. In summary, adapting existing works to meet the increasing security demands while maintaining high performance requires considerable repetitive architectural design work.
In fact, this constitutes \textbf{challenge {\ding{183}}: the absence of an efficient abstraction system that provides unified support for arbitrary pairing curves.}
Introducing such an abstraction can decouple the design process, thus enhancing both the design flexibility and performance potential. However, it also brings \textbf{challenge {\ding{184}}: hierarchical operator variants and architecture co-design complexity}.
The interdependence among software algorithms, field operator choices (arithmetic method selections), and hardware configurations creates a complex design space. Optimized operator variants shows different speed comparision results on different hardware architectures.
Existing frameworks often overlook this complexity, resulting in lost opportunities for optimization and adaptability\cite{WANG201913,ASSCC2019,bag2021flexipair,azzouzi2024novel,sakamoto2024high}.
Addressing these challenges requires a comprehensive understanding of the design space to enable effective design and improvement of PBC systems.

In this paper, we introduce \texttt{Finesse}, an agile design framework for PBC. Central to its methodology is the utilization of an abstraction system. 
This abstraction framework bridges high-level algorithms and hardware designs through clear representations and optimized mappings, providing unified support for arbitrary pairing curves in response to increasing security demands.

Through the expressive abstraction system, \Finesse's design flow has integrated a parameterized hardware architecture, an advanced compiler and a simulator to form an effective co-design cycle, enabling simple design space exploration within the complex design space.

The results from the \texttt{Finesse} framework are significant. Comprehensive evaluations show substantial performance improvements, achieving 34$\times$ throughput and 6.2$\times$ slice efficiency compared to previous flexible frameworks, and 3$\times$ throughput and 3.2$\times$ area efficiency over state-of-the-art ASIC accelerators. \texttt{Finesse} compiles code in minutes, accelerating the development cycle while ensuring that the resulting accelerators are robust and optimized for the evolving demands of modern cryptography.

To summarize, this paper makes the following key contributions:
\begin{itemize}
    \item We propose the \texttt{Finesse} design framework for pairing-based cryptography, enhancing agility in accelerator design by automating significant portions of the lengthy algorithm-to-hardware flow.
    \item We developed an abstraction system that includes IR, ISA and hardware models to support the framework.
    \item To the best of our knowledge, we are the first to incorporate \textbf{co-design mechanism} into such a framework, allowing for the effective exploration of the complex relationship within the design space.
    \item We presented a comprehensive evaluation of \texttt{Finesse} on a prototype implementation, focusing on five key aspects of the framework: accelerator efficiency, framework scalability, compilation, co-design, and agility.
\end{itemize}

\section{Background and Motivation}

\subsection{Pairing Calculation Panorama}

\begin{table}[t]
    \centering
    \caption{Symbols and notations in pairing calculation.}
    \label{tab:prelim-symbol}
    \begin{tabular}{ll}
    \toprule
    \textbf{Notation}   & \textbf{Description} \\
    \midrule
    $p$                 & size of base field \\
    $k$                 & embedding degree of curve \\
    $r$                 & pairing group order \\
    $\mathbb{F}_p$      & base prime field of size $p$ \\
    $\mathbb{F}_{p^k}$  & extension field of size $p^k$, $k$-th extension of $\mathbb{F}_p$ \\
    $E[F]$              & curve group, point coordinates in field $F$ \\
    $\mathbb{G}_1$      & pairing 1st source group ($E[\mathbb{F}_{p}]$) \\
    $\mathbb{G}_2$      & pairing 2nd source group ($E[\mathbb{F}_{p^k}]$ or $E'[\mathbb{F}_{p^{k/6}}]$) \\
    $\mathbb{G}_T$      & pairing target group ($\mathbb{F}_{p^k}$) \\
    $\ell_{Q_1, Q_2}(P)$& (tangent) line function \\
    $e(P, Q)$           & pairing function, $\mathbb{G}_1\times \mathbb{G}_2\to \mathbb{G}_T$ \\
    \texttt{M,S,A,B}    & multiplication, squaring, addition, adjunction \\
    \texttt{PA,PD}      & point addition, point doubling\\
    \bottomrule
    \end{tabular}
\end{table}

\begin{table}[t]
    \centering
    \caption{Examples of pairing-friendly curves.}
    \label{tab:curve-eg}
    \begin{tabular}{crrrrc}
    \toprule
        \multirow{2}{*}{\textbf{Curve}} & \multicolumn{4}{c}{\textbf{Param. (bit)}} & \textbf{Security} \\
    \cline{2-5}
        & $\log |t|$ & $\log p$ & $\log r$ & $k\log p$ & \textbf{(bit)} \\
    \midrule
        BN254 & 62 & 254 & 254 & 3039 & 100 \\
        BN462 & 114 & 462 & 462 & 5535 & 130 \\
        BN638 & 158 & 638 & 638 & 7647 & 153 \\
        BLS12-381 & 64 & 381 & 255 & 4569 & 123 \\
        BLS12-446 & 75 & 446 & 299 & 5352 & 130 \\
        BLS12-638 & 109 & 638 & 427 & 7656 & 148 \\
        BLS24-509 & 51 & 509 & 408 & 12202 & 192 \\
    \bottomrule
    \end{tabular}
\end{table}

Pairing is a bivariate function satisfying linearity in each of its arguments independently. Optimal Ate pairing~\cite{oatepairing2010} over elliptic curve is the de facto standard of pairing construction technique in the cryptography community, surpassing Weil pairing and Tate pairing~\cite{freeman2010taxonomy} due to its computational efficiency. The optimal Ate pairing $e(P, Q):\mathbb{G}_1\times \mathbb{G}_2\to \mathbb{G}_T$ for curve E has a complex general form, which is essentially a rational function characterized by curve-determined parameters, constructed with evaluations of the line function $\ell$ at point $P$ with respect to multiples of point $Q$. Key parameters and notations are briefly listed in~\autoref{tab:prelim-symbol}.

Pairings are defined over pairing-friendly curves to ensure computational practicality. Among these, BN~\cite{barreto2002efficient} and BLS~\cite{boneh2004short} are the most widely utilized curve families. Examples of pairing-friendly curves are characterized in~\autoref{tab:curve-eg}.

\paragraph{\textbf{Pairing Calculation}}

The optimal Ate pairing algorithm for BN and BLS curves is introduced in~\hyperref[algo:oate]{Algorithm 1}. An overview of group computation costs and their optimized alternatives summarized from~\cite{cryptoeprint:2006/471, bernstein2007faster} is presented in~\autoref{tab:prelim-cost}. The algorithm consists of two main components: the Miller loop and final exponentiation, which account for approximately 40\% and 60\% of the overall computation cost, respectively.

The Miller loop involves iterative calculation of values of the Miller function through point doublings and additions, as well as evaluations of line and tangent functions. Common optimization strategies for the Miller loop include the use of non-adjacent forms and the integration of point operations with line operations.

The final exponentiation transforms Miller function values into canonical form, ensuring consistency by resolving coset equivalence relation. Key optimizations include Frobenius-based techniques for the easy part and decomposition and reuse strategies for the hard part, as demonstrated by~\cite{devegili2007implementing} and~\cite{fuentes2011faster}. Additionally, operations within the cyclotomic subfield are optimized to further improve efficiency.

\begin{table}[t]
    \centering
    \caption{Costs in pairing calculation.}
    \label{tab:prelim-cost}
    \begin{tabular}{llll}
    \toprule
        \textbf{Group} & \textbf{Storage} & \multicolumn{2}{l}{\textbf{Operation Costs}}\\
    \midrule
    $\mathbb{F}_p$ & $(\log p)$ bit & $\texttt{M}_1,\texttt{S}_1\in \mathcal{O}(\log^{1.58} p)$ & $\texttt{A}_1,\texttt{B}_1\in \mathcal{O}(\log p)$ \\
    $\mathbb{F}_{p^{2d}}$ & $2\,\mathbb{F}_{p^{d}}$ & \multicolumn{2}{l}{$\texttt{M}_{2d}=4\texttt{M}_{d}\,2\texttt{A}_{d}\,1\texttt{B}_{d}$ or $3\texttt{M}_{d}\,5\texttt{A}_{d}\,1\texttt{B}_{d}$ or $\dots$} \\
    $\mathbb{F}_{p^{3d}}$ & $3\,\mathbb{F}_{p^{d}}$ & \multicolumn{2}{l}{$\texttt{M}_{3d}=9\texttt{M}_{d}\,6\texttt{A}_{d}\,2\texttt{B}_{d}$ or $5\texttt{M}_{d}\,33\texttt{A}_{d}\,2\texttt{B}_{d}$ or $\dots$} \\
    \midrule
    \multirow{2}{*}{$E[\mathbb{F}_{p^d}]$} & \multirow{2}{*}{$3\,\mathbb{F}_{p^{d}}$} & \multicolumn{2}{l}{$\texttt{PA}_{d}=11\texttt{M}_{d}\,5\texttt{S}_{d}\,13\texttt{A}_{d}$ or $12\texttt{M}_{d}\,25\texttt{A}_{d}$ or $\dots$} \\
    & & \multicolumn{2}{c}{$\texttt{PD}_{d}=2\texttt{M}_{d}\,5\texttt{S}_{d}\,11\texttt{A}_{d}$ or $5\texttt{M}_{d}\,6\texttt{S}_{d}\,11\texttt{A}_{d}$ or $\dots$} \\
    \bottomrule
    \end{tabular}
\end{table}

\begin{algorithm}[t]
    \caption{Optimal Ate Pairing (BN/BLS)\label{algo:oate}}
    \KwIn{$P \in \mathbb{G}_1, Q \in \mathbb{G}_2$}
    \KwOut{$e(P,Q) \in \mathbb{G}_T$}
    \SetKw{KwDownTo}{downto}
    \SetKw{KwIs}{is}
    \If{curve family \KwIs \rm{BN}}{
        $u \gets 6t + 2$
    }
    \ElseIf{curve family \KwIs \rm{BLS}}{
        $u \gets t$
    }
    $(T, f) \gets (Q, 1)$\\
    \For(\tcc*[f]{Miller Loop}){$i \gets \lfloor \log u \rfloor$ \KwDownTo $0$ }{
        $(T,f) \gets ([2]T, f^{2} \cdot \ell_{T}(P))$\\
        \If{$u[i]=1$}{
            $(T,f) \gets (T + Q, f \cdot \ell_{T,Q}(P))$
        }
    }
    \If{curve family \KwIs \rm{BN}}{
        $Q_1 \gets \mathrm{frob}(Q)$ \\
        $Q_2 \gets -\mathrm{frob}(Q_1)$ \\
        $(T,f) \gets (T+Q_1, f \cdot \ell_{T,Q_1}(P))$ \\
        $(T,f) \gets (T+Q_2, f \cdot \ell_{T,Q_2}(P))$
    }
    $f \gets f^{(p^{k}-1)/r}$ \tcc*[f]{Final Exponentiation} \\
    \Return $f$
\end{algorithm}

The security of a given pairing diminishes as cryptographic attack algorithms evolve, while computational power continues to increase. To maintain a suitable security level (128/192/... bits), applications are adopting curves with progressively wider bit-widths. Additionally, to preserve the balance between the hardness of the FFDLP/ECDLP while keeping bit-width in a reasonable range, the embedding degree must also increase~\cite{freeman2010taxonomy}. Although different curve families share commonalities in terms of extension fields, twist curves, and the fundamental algorithmic framework, their computational details vary significantly.

\paragraph{\textbf{Insights}}
\label{para:prelim-insight}
Pairing has applications in various cryptographic fields. In the Groth16~\cite{groth2016size} zero-knowledge proof system, pairing is used to verify the correctness of a proof by checking whether an equation involving a bilinear relation holds. Pairing significantly reduces the proof size, making verification more efficient.

The key aspects of pairing computation lie in bit-width, extension field arithmetic and reduction costs dominated by embedding degree and twist degree, as well as the optimization methods applied to the final exponentiation. As these requirements evolve, pairing computations tend to introduce wider multiplication widths, more complex control flow, greater pressure on memory access patterns.

\subsection{Motivations}

\paragraph{\textbf{\scalebox{1.1}{\ding{182}} Mitigate Re-engineering Costs to Keep Pace with Growing Cryptographic Demands}}
Designing accelerators for bilinear pairings has long been a complex and resource-intensive endeavor. The inherently multi-layered nature of pairing computations often makes structural approaches a natural choice, as it encourages designers to map algorithmic hierarchies directly into circuit hierarchies~\cite{xie2022high, WANG201913}. Alternatively, some works resort to manually transcribing algorithms into microcodes~\cite{sakamoto2024high}, handling each operator at a fine-grained level. While both approaches can achieve functional correctness, they commonly require extensive human intervention in tasks such as operator decomposition, control signal generation and operation scheduling, resulting in substantial design costs and extended development cycles. In summary, expanding the scope of these works to support a broader range of pairing curves or updating them would present significant challenges.

\begin{figure}[t]
    \centering
    \includegraphics[width=\linewidth]{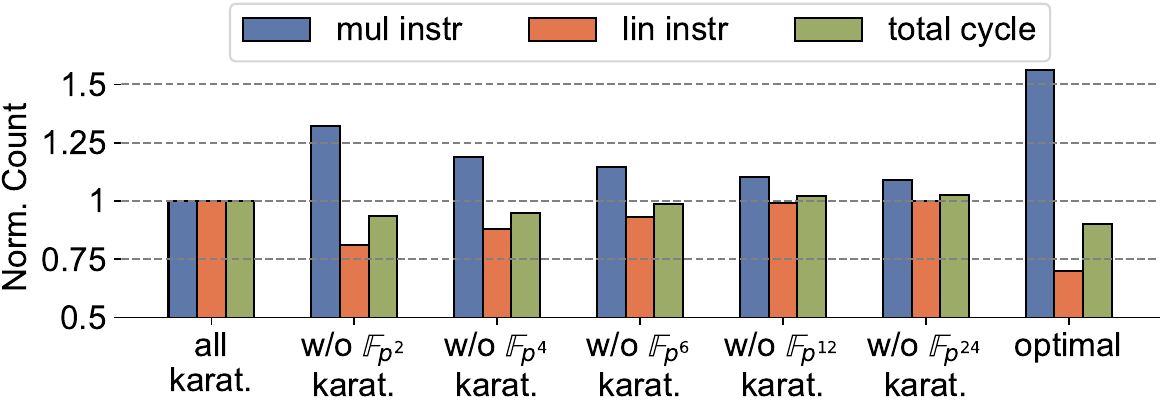}
    \caption{Comparison of operator-level optimization combinations. Curve: BLS24-509, Algo: O-Ate.}
    \label{fig:op-comb-exp}
    \Description{A comparative pie chart showing normalized counts of \texttt{mul}, \texttt{lin}, and total instructions, along with total cycle count, across several configurations of the "karat" strategy with varying numbers of components removed (p2 to p24) and an optimal baseline. Each slice represents one configuration, comparing its efficiency in terms of these metrics.}
\end{figure}

\begin{figure*}[t]
    \centering
    \includegraphics[width=1\linewidth]{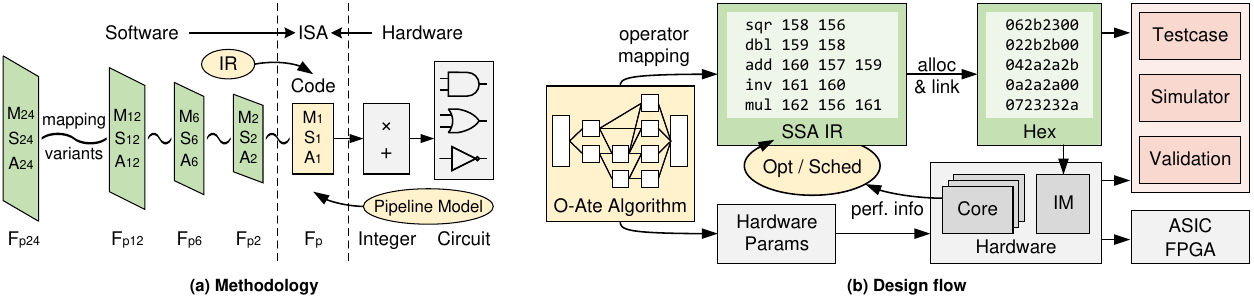}
    \caption{Overview of \texttt{Finesse} design framework.}
    \Description{A diagram outlining a framework for mapping high-level O-Ate pairing algorithms onto hardware via an intermediate representation (SSA IR), with scheduling and optimization. It connects algorithmic operations (e.g., \texttt{add}, \texttt{mul}, \texttt{inv}) to hardware instructions, flowing through simulation, validation on ASIC/FPGA, and performance testing. The design supports various field structures (Fp, Fp2, Fp6, Fp12, Fp24) and corresponding operations (e.g., M1, M2, ..., M24).}
    \label{fig:sw-arch}
\end{figure*}

Furthermore, the structural differences across pairing families and curve types highlight the need for greater agility in the design process. Agility enables designers to adapt quickly to diverse requirements, reducing the overhead of manual re-engineering, and accelerating the design cycle. By fostering faster prototyping and iterative feedback, \textbf{an agile framework addresses the inefficiencies of traditional methods and empowers pairing accelerator designs to keep pace with evolving cryptographic demands}.

\paragraph{\textbf{\scalebox{1.1}{\ding{183}} Propose an Efficient Abstraction to Bring Performance and Flexibility}}
In light of growing needs, recent works have stumbled around the threshold of flexibility. FlexiPair~\cite{bag2021flexipair} is a flexible pairing framework that aims to provide flexibility for edge devices at the expense of performance potential, revealing significant limitations in its approach. Its framework employs a fixed hardware architecture, which lacks hardware abstraction and co-design capabilities, resulting in limited extensibility. Performance improvements are constrained primarily by bottlenecks in memory and ALU operations. Furthermore, the optimization strategies employed are insufficient, neglecting the potential for software compilation optimizations and software-hardware co-optimization.

On the other side, traditional accelerator designs have completely ruled out the option for flexibility. The current SOTA accelerator on ASIC~\cite{ASSCC2019} serves as a prime example of this limitation. Its customized computational layers, particularly the ALU specialized for $\mathbb{F}_{p^2}$, are not adaptable to non-$\mathbb{F}_{p^2}$ curves, thereby limiting its versatility. Additionally, their optimization methodology for customized pipeline structure focuses solely on a narrow scheduling window, mapping $\mathbb{F}_{p^{12}}$ to $\mathbb{F}_{p^2}$, which ignored the potential of global optimization.

To achieve the coexistence of high performance and flexibility, an effective abstraction system is needed to fully cover the design process.
Such an abstraction should embrace both algorithmic and architectural possibilities, supporting various curves, operators, and hardware components in a compatible and extensible manner. To summarize, \textbf{a compatible and extensible abstraction is the common soil for achieving both flexibility and high-performance in overall architecture}.

\paragraph{\textbf{\scalebox{1.1}{\ding{184}} Exploring the Complex Design Space for Optimization}}
\label{para:motivation-dse}
We begin by examining a small experiment related to Karatsuba optimization over finite fields. Karatsuba is an optimization technique aimed at reducing the number of multiplications, at the cost of increasing the number of linear operations. While this method proves effective on platforms like CPUs, its benefits may not translate as clearly to hardware accelerators.

This discrepancy arises because accelerators typically access memory with a width that directly matches the base field bitwidth. Both linear and multiplication operations exert the same memory bandwidth pressure, but linear operations perform less computation per memory access. Consequently, linear operations result in lower computational throughput per memory load. Furthermore, on single-issue architectures, when both types of operations occupy one full cycle in the pipeline, the increased number of linear operations exacerbates this imbalance in the instruction issue queue.

However, for higher-degree fields (i.e. $\mathbb{F}_{p^{12}}$ or $\mathbb{F}_{p^{24}}$), the advantage of Karatsuba method cannot be ignored. High-level multiplications are decomposed into more $\mathbb{F}_p$ instructions ($k^2$ or $k^{1.585}$, while linear operations only break down into $k$ $\mathbb{F}_p$ instructions. As shown in~\autoref{fig:op-comb-exp}, we conducted a validation experiment on a basic single-issue architecture. By disabling Karatsuba-like optimizations in the $\mathbb{F}_{p^2}$ or $\mathbb{F}_{p^4}$ operators, we observed a reduction in the overall cycle count, compared to using optimization on all levels. 

The impact of algorithmic optimizations varies significantly depending on the underlying hardware design, demonstrating that effective DSE (design space exploration) is crucial for uncovering configurations that maximize performance while efficiently utilizing hardware resources.

The design space for pairing accelerators is shaped by factors like operator variants, IP availability and hardware data path structures, requiring careful coordination between components and architecture. Diverse and often conflicting goals---such as optimizing area, throughput, or area/delay trade-offs---add complexity to the exploration. In this context, \textbf{DSE and co-design frameworks are crucial, enabling systematic evaluation of design options and hardware-software co-optimization to unleash the potentials of pairing accelerators}.

\section{Finesse Framework}
\subsection{Framework Overview}
 
As sketched in~\autoref{fig:sw-arch}, \texttt{Finesse} is an agile full-process design framework for pairing-based cryptography, offering a comprehensive and extensible system that spans from high-level algorithm description to low-level hardware models, providing direct validation and deployment across various cryptographic curves. \texttt{Finesse} comes with a fully functional implementation, providing out-of-the-box support for popular curves. 

\paragraph{\textbf{Methodology}} 

Pairing accelerator design is inherently a multi-layered problem, spanning from high-level finite field operators (e.g. $\texttt{M}_{24}$) and point operators (e.g. $\texttt{PA}_{4}$) down to low-level circuit design. This hierarchical structure introduces significant challenges in mapping algorithms efficiently to hardware. To address this, \texttt{Finesse} adopts a co-design methodology that strategically partitions software and hardware layers while providing abstractions tailored to each level of the computation.

The boundary between software and hardware is defined by an instruction set architecture (ISA) abstraction. This abstraction standardizes interactions between algorithmic logic and hardware, serving as a modular interface. Above the ISA, high-level computations are represented through an intermediate representation (IR). The IR enables software optimizations, such as instruction scheduling and dependency analysis, while maintaining flexibility for different hardware backends. Below the ISA, hardware implementations rely on pipeline models to describe its behavior, including latency, memory and instruction issue parameters. These models guide the hardware realization process, ensuring alignment with upstream abstractions while permitting design variations. By integrating these abstractions, \texttt{Finesse} establishes a systematic framework for software/hardware co-design, supporting pairing computations across diverse configurations.

\paragraph{\textbf{Design Flow}}

The design flow of \texttt{Finesse} is organized into a set of modular components, each addressing specific aspects of pairing accelerator development. The hardware and simulator part uses abstraction to define and implement pipeline behavior, enabling rapid exploration of configurations. The compiler and optimization part maps high-level operators into IR, supporting co-design between software and hardware. It applies scheduling and data flow optimizations while enabling basic DSE, which systematically evaluates performance and resource trade-offs. The validation part verifies correctness and performance on simulators and prototype platforms (ASIC and FPGA), generating feedback for iterative design refinement. \texttt{Finesse} also offers a basic operator kit for quickly porting new curves or pairing algorithms into the framework. These components work together to streamline the development process and support a wide range of pairing accelerator configurations.

\subsection{Abstraction}

\label{para:abs-description}

Abstraction in the \texttt{Finesse} framework defines clear interfaces between software and hardware layers, enabling efficient mapping of high-level algorithms to hardware. It utilizes intermediate representations (IR), instruction set architectures (ISA), and hardware models to optimize and flexibly represent both software and hardware components. This structured approach ensures interoperability, scalability, and streamlined co-design, supporting diverse pairing accelerator configurations.

\begin{table}[t]
    \centering
    \caption{IR operations and their supported argument data type. \texttt{fp-like} = \texttt{fp} or \texttt{fpd}, \texttt{ep-like} = \texttt{ep} or \texttt{epd}. Adjoined element refers to the element adjoined to the base field for defining the extension field.}
    \label{tab:abs-ir}
    \begin{tabular}{rll}
    \toprule
    \textbf{IR Op.} & \textbf{Description} & \textbf{Supp. Arg. Type} \\
    \midrule
    \texttt{add/sub} & field addition/subtraction & \texttt{(fp-like, fp-like)}\\
    \texttt{muli} & field scalar multiplication & \texttt{(int, fp-like)}\\
    \texttt{mul} & field multiplication & \texttt{(fp-like, fp-like)}\\
    \texttt{sqr} & field squaring & \texttt{fp-like}\\
    \texttt{exp} & field exponentiation & \texttt{(fp-like, int)}\\
    \texttt{adj} & multiply by adjoined el. & \texttt{fpd}\\
    \texttt{conj} & conjugate to adjoined el. & \texttt{fpd}\\
    \texttt{frob} & Frobenius endomorphism & \texttt{(fp-like, int)}\\
    \texttt{padd} & curve point addition & \texttt{ep-like}\\
    \texttt{pmul} & curve scalar multiplication & \texttt{(int, ep-like)}\\
    \bottomrule
    \end{tabular}
\end{table}

\begin{figure}[t]
    \centering
    \includegraphics[width=\columnwidth]{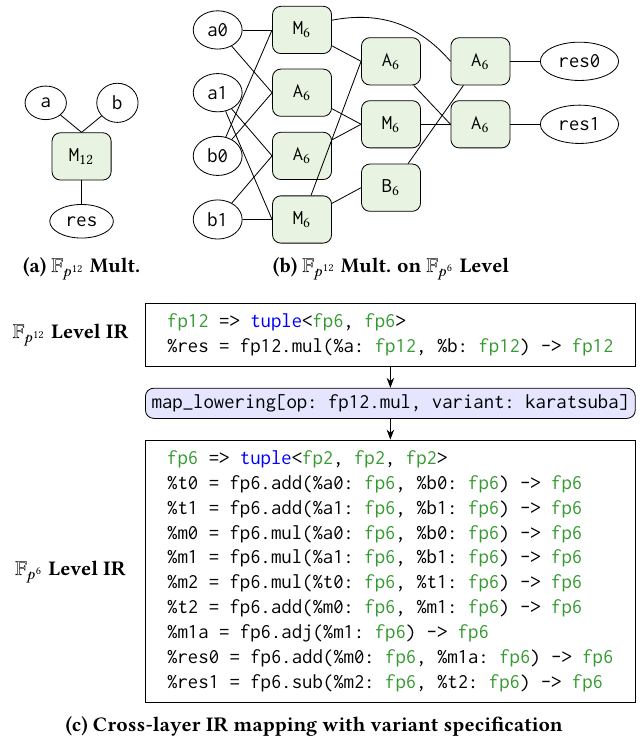}
    \caption{Example of \texttt{Finesse}'s operator mapping through IR abstraction.}
    \label{fig:abs-ir-eg}
    \Description{A three-part illustration of a modular multiplication in Fp12, showing (a) the high-level structure of Fp12 multiplication, (b) its decomposition into Fp6 operations using Karatsuba, and (c) the corresponding IR mapping across Fp12 and Fp6 levels. The IR expresses the operation as structured tuples and typed SSA-form instructions, including explicit lowering rules and intermediate temporaries.}
\end{figure}

\paragraph{\textbf{Challenges}}
Main consideration in abstraction design is focused on compatibility and extensibility. On the algorithm side, abstraction needs to support the family/curve/operator triplet by capturing commonalities and covering diverse PBC primitives. We opted to keep it as simple as possible to preserve extensibility towards broader fields in cryptography. On the hardware side, \Finesse~needs to be compatible with a series of architectural candidates, a wide portfolio of ALUs and memory units ranging from open-source free designs, self-made designs and available proprietary IP cores. Also \Finesse~should be able to evolve and support novel datapaths that provides better parallelism in the future. This can be achieved through ISA-level extensions and improvements on instruction selection strategies. Facing these challenges, \Finesse~has chosen to define abstractions carefully and move complexities to above ISA level rather than sub-ISA level, solving the problem mainly at compile time.

\paragraph{\textbf{Abstraction Design}}

\Finesse~IR is mainly focused on expressing calculation on algebraic objects. Custom data types include: \texttt{fp} $\to$ $\mathbb{F}_p$, \texttt{fpd} $\to$ $\mathbb{F}_{p^d}$, \texttt{ep} $\to$ $E[\mathbb{F}_{p}]$, \texttt{epd} $\to$ $E[\mathbb{F}_{p^d}]$, where $d$ denotes the dimension of the extension field relative to base field $\mathbb{F}_p$. Necessary parameters (refer to~\autoref{tab:prelim-symbol} for a list) determining the field structures and curve structures is incorporated as attributes to the IR. \autoref{tab:abs-ir} gives a list of defined operations on these objects. As a simplification, operations between \texttt{fp-like} objects or \texttt{ep-like} objects requires divisibility on their dimension parameters \texttt{d} (or else an efficient homomorphism would be required, which is possible, but over-complicates the abstraction system).

\autoref{fig:abs-ir-eg} illustrates an example of \texttt{Finesse}'s abstraction system, showcasing the transformation of a high-level $\mathbb{F}_{p^{12}}$ multiplication operation into a lower-level $\mathbb{F}_{p^{6}}$ representation through IR. The upper part of the figure depicts the $\mathbb{F}_{p^{6}}$ operation expressed using abstract operators, while the lower part demonstrates the detailed decomposition into $\mathbb{F}_{p^{6}}$ operators, using the Karatsuba variant. Following this approach the framework bridges high-level algorithmic constructs with hardware-aware granular operations, ensuring efficient and modular design.

\Finesse~defines a simple RISC-flavor $\mathbb{F}_p$-level ISA with VLIW extension. Machine operations include: linear operations (\texttt{NEG}, \texttt{DBL}, \texttt{TPL}, \texttt{ADD}, \texttt{SUB}), multiplicative operations (\texttt{SQR}, \texttt{MUL}), inverse operation (\texttt{INV}), miscellaneous (\texttt{NOP}, \texttt{CVT}, \texttt{ICV}). \Finesse~performs computation in its dedicated on-chip/on-fabric register banks, thus all operands are registers. \texttt{CVT} and \texttt{ICV} operations are designed for post/pre I/O data format conversions. In VLIW extension, multiple operations can be packed into a single ``wide instruction'', or ``issue slot'', to enable explicit ILP.

\Finesse's hardware model describes hardware resource information and instruction itineraries. These include: number of linear ALUs, number of register banks and register quota per bank, simultaneous read/write capacity for each bank, presence of write back ring buffers for register banks, maximum operations allowed in a single VLIW instruction (similar to that of issue width in multi-issue architecture), delay info and resource consumption for each operation (itineraries). Currently \Finesse~asserts a few reasonable constraints on the model: at most 1 \texttt{mmul} ALU per core, at least many register banks as VLIW width, at least 2 reads + 1 writes per bank per cycle, existence of write back ring buffers on VLIW architectures (width $\ge 2$).

\paragraph{\textbf{Abstraction Overhead}}
In \Finesse, abstraction above ISA layer does not involve complex programming paradigm, and is essentially zero-cost, i.e. no additional control information stored at runtime. When decomposed onto $\mathbb{F}_p$ level operations, higher-level constuct info can be dropped, and constants needed in lowering mapping can fit in a small table to be fetched at runtime. Assertions in and below ISA inevitably hide some possible optimization opportunity, but this tradeoff is worthy in a systematic approach for agile co-design.

\subsection{Hardware Architecture}

\begin{figure}[t]
    \centering
    \includegraphics[width=\linewidth]{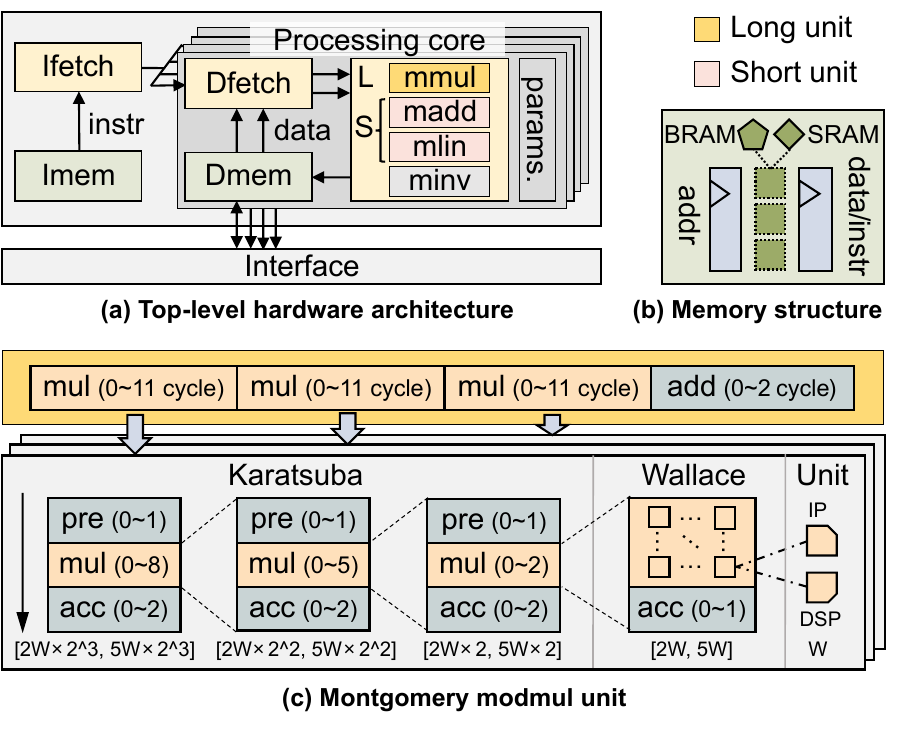}
    \caption{Hardware architecture of \texttt{Finesse} framework.}
    \label{fig:hw-arch}
\end{figure}

With abstraction decoupling the software and hardware, the hardware only needs to focus on implementing the pipeline model architecture, which is designed to support operations over $\mathbb{F}_p$. \autoref{fig:hw-arch} provides an overview of the basic hardware architecture supported by our \texttt{Finesse} framework. As shown in~\autoref{fig:hw-arch}(a), the architecture is built on a pipeline structure, consisting of instruction memory/fetch units, and one or more processing cores, each of which includes data memory/fetch units and an ALU. The ALU features four modular arithmetic units designed for $\mathbb{F}_p$ operations, including modular multiplication, modular addition, modular doubling (in the mlin unit), and modular inversion, among others. 

We employed the Jacobian coordinate system to implement the pairing algorithm, which requires modular inversion (performed in the minv unit) only once. Consequently, the relatively complex minv unit is designed using an iterative structure, while all other computation units adopt a fully pipelined structure. In our prototype, modular multiplication is treated as a \texttt{Long} pipeline unit, whereas other linear units are considered \texttt{Short} pipeline units.

\paragraph{\textbf{Parameterization}}

Through parameterized design, the architecture can be adapted to diverse application requirements and platform constraints, while providing a foundation for performance optimization. The primary parameters include curve constants, data width, memory configuration, the number of parallel cores, base unit mapping to various IP portfolios, and the pipeline depth of computation units. Key benefits of parameterization include:

\begin{itemize}
    \item \textbf{Adaptability} to different curve scenarios via parameterization of curve constants, data width, and memory size.
    \item \textbf{Flexibility} in adapting to varying throughput requirements through adjustment of the parallel core count.
    \item \textbf{Platform independence} enabled by the mapping of base units to different deployment platforms (e.g. ASIC, FPGA).
    \item ALU family co-design optimization is supported by parametrizing the \textbf{pipeline depth} of key computational units.
\end{itemize}

\paragraph{\textbf{Multi-core}}

\begin{figure}[t]
    \centering
    \includegraphics[width=1.0\linewidth]{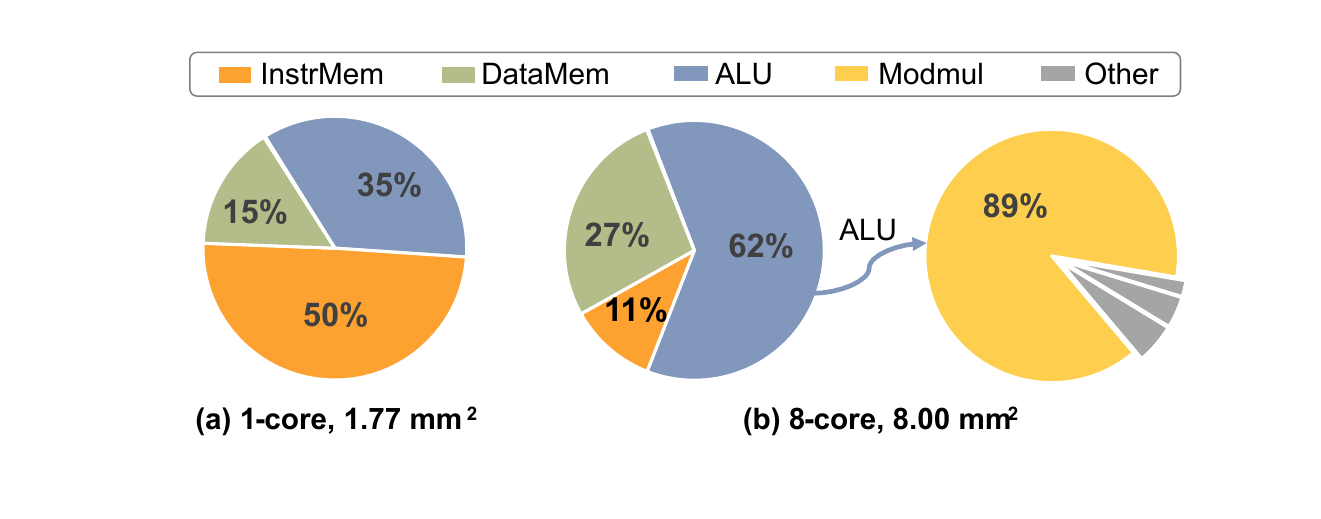}
    \caption{Hardware area breakdown. Hardware model: \texttt{Long} = 38cy, \texttt{Short} = 8cy, \texttt{2R1W}/cy. Curve: BN254N, Algo: O-Ate.}
    \label{fig:hw-area}
    \Description{Explained in the text.}
\end{figure}

As shown in~\autoref{fig:hw-area}, instruction memory occupies 50\% of the area in a single-core design. Analysis reveals that, for pairing computations on the same curve, the operations are identical. This consistency allows us to replicate multiple data memory and ALU while utilizing a shared instruction memory. \autoref{fig:hw-area}(b) presents the area breakdown of an 8-core design, where instruction memory accounts for only 11\% of the total area. This reduction highlights better area utilization, as the total area increases by 4.5 times while achieving an 8-fold improvement in overall throughput, resulting in a 77\% gain in area efficiency. In fact, in our architectural design, the number of parallel cores and area efficiency aligns with Amdahl's law.

Indeed, this parallel approach aligns with the SIMT architecture, which meets the demands of high-throughput applications and further enhances area efficiency.

\paragraph{\textbf{Optimizations}}

\label{para:opt}
It is important to note that the area and timing are directly influenced by parameterization. Firstly, the design must ensure that the storage and computation units support different data width, and the computation units support pipeline depth parameterization. Secondly, from a co-design perspective, the pipeline depth of the ALU itself affects the overall timing performance (refer to \autoref{fig:eval-dse-alu} in the below).
To enhance the efficiency of pairing operations, timing and area optimizations were applied to parameterized storage and computation units.

For storage units, the design enables arbitrary bitwidth and depth by automatically combining small basic memory units into larger configurations. As depicted in~\autoref{fig:hw-arch}(b), to reduce path delay caused by combining smaller memory units into larger ones, registers are placed before and after the memory, creating a three-stage pipeline for read/write operations. 
The attributes of the basic memory block are fixed by IP vendors. So the storage area (i.e., the number of basic memory block used) is more dependent on the binary size the compiler generates (for IMem) and the maximum number of active registers (for DMem).
Meanwhile, the basic memory units can be mapped to specific platform primitives.

As illustrated in~\autoref{fig:hw-area}(b), Modular multiplication (mmul) unit is the core of the ALU, which accounts for 89\% of the ALU's area. Our primary objective is to optimize mmul while achieving parameterization of bitwidth and depth. In terms of timing, we adopt a deeply pipelined design, decomposing modular multiplication into multiple stages to achieve high throughput. Regarding area, we reduce the number of multiplications by leveraging the Karatsuba algorithm. To facilitate optimization within a parameterized framework, we design a hierarchical modular multiplication module, as illustrated in~\autoref{fig:hw-arch}(c). At the basic unit layer, the multiplication width $W$ determines the critical delay path, as it is directly mapped to FPGA DSP blocks or ASIC multiplier IPs. To enable efficient module partitioning, we encapsulated $2W$ to $5W$ bit multiplier modules using the Wallace tree algorithm based on the basic unit. Further optimized by recursively applying the integer Karatsuba multiplication algorithm $n$ times, the structure covers a range from $2W\times2^n$ to $5W\times2^n$, which effectively reduces the multiplier's area. For instance, with $W = 16$ and $n = 3$, the proposed approach achieves an approximate 40\% reduction in area compared to naive multiplication.

\subsection{Simulator}
\label{para:arch-simulator}

The simulator empowers both software and hardware validation flows. On the algorithmic side, to enable validation of post-compile code execution trace, we have implemented a single-cycle functional simulator capable of executing SSA instructions. The correctness verification is accomplished through cross-validation of computational results against established cryptographic libraries such as MCL~\cite{herumi_mcl}, MIRACL~\cite{miracl_miracl}, and RELIC~\cite{relic-toolkit}.

At the hardware level, we have implemented a cycle-accurate simulator consistent with the RTL behavior based on the pipeline model to simulate instruction delay and data dependence. Within our framework, this simulation platform serves as a experimental infrastructure, providing data references for works like compiler affinity optimization and design space exploration.

\subsection{Compilation Techniques}
\label{para:compile-tech}

\paragraph{\textbf{Compilation Pipeline}}
In contrast to modern general-purpose compilers, \texttt{Finesse} uses a shorter compilation pipeline, following the execution order:

\begin{itemize}
\item \textbf{CodeGen}: Simplified leveraging algorithmic characteristics. In the optimal Ate pairing algorithm, both the Miller loop and final exponentiation have fixed loop parameters based on the underlying curve, allowing convenient loop unrolling and restructuring into a single basic block. The code generation task is completed by mapping the algorithm into IR, with respect to operator variants.
\item \textbf{IROpt}: Standard SSA-based data flow optimizations, with additional assumptions from finite field arithmetic. Specifically, constant propagation with Frobenius constant tables, strength reduction, dead code elimination and global value numbering using commutativity on finite fields.
\item \textbf{BankAlloc}: Assigns values to register banks. A simple residual assignment strategy serves as an effective baseline.
\item \textbf{PackSched}: Schedules and packs operations into issue slots. Supports both single issue architectures and VLIW architectures.
\item \textbf{RegAlloc}: Sequential register allocation within banks, based on liveness analysis.
\item \textbf{ASM}: Translates IR into hardware-defined instruction encoding.
\item \textbf{Link}: Consolidates various basic blocks into a single binary and resolves entry address offsets.
\end{itemize}

\paragraph{\textbf{Scheduling Strategies}}

\begin{figure}[t]
    \centering
    \includegraphics[width=\linewidth]{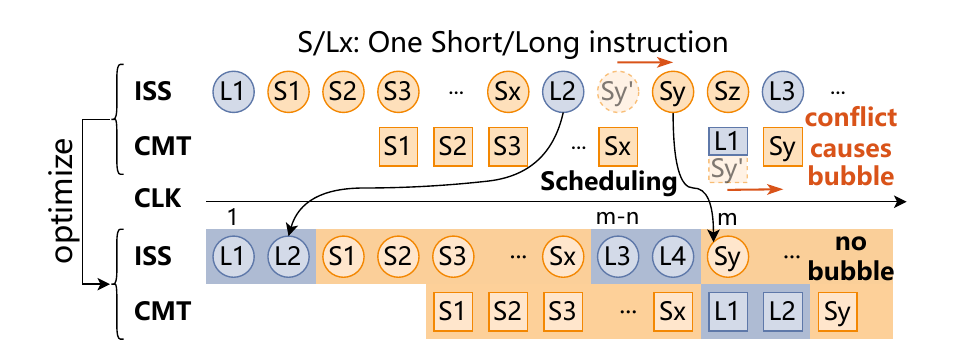}
    \caption{Illustrated example of instruction issue slot affinity optimization. \texttt{Long} = $m$cy, \texttt{Short} = $n$cy. \texttt{L1} and \texttt{Sy'} are $(m-n)$cy apart, causing latter to stall until \texttt{Sy}. Setting issue slot affinity helps compiler to avoid this issue.}
    \Description[]{Fully described in the caption.}
    \label{fig:instr-aff}
\end{figure}

\begin{algorithm}[t]
    \caption{Operation Packing and Scheduling}
    \label{algo:aff-sch}
    \KwIn{IR code $\mathcal{P}$, in SSA form}
    \KwOut{A valid schedule $\mathcal{S}$, satisfying HW constraints}
    \SetKw{KwDownTo}{downto}
    \SetKw{KwWhere}{where}
    \SetKw{KwBreak}{break}
    \SetKwProg{KwFn}{Function}{}{}
    \SetKw{KwIn}{in}
    \SetKw{KwIs}{is}
    $\mathcal{S}\gets [\;]$\\
    $deps \gets \{\;\}$ \\
    $trigger\gets \{\;\} $ \\
    $queue\gets [\;]$

    \KwFn{\emph{\textsc{solveMaxValidInstrPack}($now$)}}{
        \tcp*[h]{state: (maxInst, candInst)}\\
        $f\gets {\{\varnothing: (0, [\;])\}}$\\
        $bestState\gets \{\;\}$ \\
        $issueQueue\gets \text{collect readied instr from } queue \text{ at } now$ \\
        \For{$instr$ \KwIn $\operatorname{sortByAffinity}(issueQueue)$}{
            $state_1 \gets \text{compress itinerary of }instr\text{ to DP state}$ \\
            \For{$state_2 \in f$}{
                \If{$state_1$ does not contradicts $state_2$}{
                    $new \gets state_1 \cup state_2$ \\  
                    update $f[new]$ if $f[state_2]+instr$ is better\\
                    update $bestState$ by $f[new]$
                }
            }
        }
        \Return $f[bestState].\!\operatorname{candInst}$
    }

    \For{$instr$ \KwIn $\mathcal{P}$} {
        $deps[instr] \gets\operatorname{countDependency}(instr)$ \\
        \If{$\operatorname{isConstOp}(instr)$} {
            $\mathcal{S}.\!\operatorname{append}(instr)$
        }
        \Else {
            $trigger[instr.operand].\!\operatorname{append}(instr)$
        } 
    } 
    $issueTime\gets 0$ \\
    $n \gets \operatorname{len}(\mathcal{P})-\operatorname{\#ConstOp}$ \\
    \While{$n$ > 0} {
        $instrs \gets \operatorname{solveMaxValidInstrPack}(issueTime)$ \\
        $\mathcal{S}\gets\mathcal{S}+instrs$ \\
        \For{$instr$ \KwIn $instrs$}{
            \For{$inst$ \KwIn $trigger[instr]$}{
                $deps[inst] \gets deps[inst] - 1$ \\
                \If{$deps[inst]=0$}{
                    $queue\gets queue+[instr]$
                }
            }
            $queue\gets \operatorname{delete}(queue, instr)$
        }
        $n\gets n-\operatorname{len}(instrs)$ \\
        $issueTime\gets issueTime+1$
    }
    \Return $\mathcal{S}$ \\    
\end{algorithm}

\texttt{Finesse} supports scheduling strategies for both single-issue and VLIW architectures. Scheduling is performed in a single pass immediately after BankAlloc, where values are assigned to register banks without being mapped to specific registers yet. \texttt{Finesse} uses a top-down scheduling algorithm with affinity-based selection order and constraint solving for register bank read/write constraints.

The core innovation of \texttt{Finesse} scheduling strategy lies in its \textit{issue slot affinity optimization}. \autoref{fig:instr-aff} provides a conceptual example of this approach. From a hardware abstraction perspective, R/W operations in memory blocks are constrained by certain limitations, and conflicts can arise when a \texttt{Long} instruction is issued followed by a \texttt{Short} instruction after a certain delay. To address this, \texttt{Finesse} partitions the issue slots into periodic intervals based on the difference in cycle counts between \texttt{Long} and \texttt{Short} instructions. Within each interval, specific positions are assigned \texttt{Long} instruction affinity, while the rest are assigned \texttt{Short} instruction affinity. Following notation of~\autoref{fig:instr-aff}, we can formulate the affinity for issue slot at cycle $T$ as
$$
\text{Affinity}(T) \coloneqq \dfrac{T\bmod(m-n)}{m-n}\le \dfrac{\#\,\texttt{LongInstr}}{\#\,\texttt{Instr}}+\beta,
$$
where $\beta$ is a tunable parameter, and $\text{Affinity}(T)=\texttt{True}$ implies \texttt{Long} affinity, \texttt{Short} affinity otherwise. 

\texttt{Finesse} employs a constraint solving process combining data dependency, issue slot affinity and instruction itineraries to produce a correct and efficient schedule for the IR code. \hyperref[algo:aff-sch]{Algorithm 2} gives a detailed formulation of this algorithm. It starts by scanning top-down on DAG structure, and on each cycle $T$, a table of candidate instructions are drawn from those that are ready to be issued at current cycle, following the order determined by $\text{Affinity}(T)$. We use dynamic programming to check for maximum combination of operations into a single issue slot without violation of R/W constraints.

\begin{table}[t]
    \centering
    \caption{Examples of operator variants for key extension fields in curve BLS24-509.}
    \label{tab:dse-opv}
    \begin{tabular}{lll}
    \toprule
    \textbf{Group} & \textbf{Op.} & \textbf{Variants} \\
    \midrule
    \multirow{2}{*}{$\mathbb{F}_{p^{6}}$}   & $\texttt{M}_{6}$ & Karatsuba, Schoolbook \\
                                    & $\texttt{S}_{6}$ & CH-SQR\{1,2,3\}~\cite{chung2007asymmetric}, Complex, Schoolbook \\
    \midrule
    \multirow{2}{*}{$\mathbb{F}_{p^{12}}$}  & $\texttt{M}_{12}$ & Karatsuba, Schoolbook \\
                                    & $\texttt{S}_{12}$ & Complex, Schoolbook \\
    \midrule
    $\mathbb{G}_2$  & $\texttt{PA}_{4},\texttt{PD}_{4}$ & Jacobian, Projective \\
    \bottomrule
    \end{tabular}
\end{table}

\subsection{Design Space Exploration}

As a first step towards comprehensive DSE, \texttt{Finesse} concentrates on solving the problem in a pivotal subspace of the general design space. The design space in our framework is characterized by two key elements: \textit{operator variants combination} and \textit{hardware model}, both of which significantly influence the performance and resource efficiency of the accelerator.

\begin{table*}[t]
    \centering
    \caption{Evaluation comparison on the BN254/BN256 curves. Hardware model: \texttt{Long} = 38cy, \texttt{Short} = 8cy, \texttt{2R1W}/cy. Algo: O-Ate.}
    \label{tab:prf-total}
    \begin{minipage}{2\columnwidth}
    \begin{tabular}{llrrrrrr}
    \toprule
    \textbf{Work} & \textbf{Platform} & \textbf{Frequency} & \textbf{\#Cycle} & \textbf{Latency} & \textbf{Util./Area} & \textbf{Throughput} & \textbf{Throughput/Area} \\
    \midrule
    ~\cite{bag2021flexipair} & FPGA Virtex-7 & \SI{188.5}{\mega\hertz} & 2552k & \SI{14.14}{\milli\second} & 2506 Slices & 70.7 ops & 0.028 ops/Slice \\
    Ours (1-core) & FPGA Virtex-7 & \SI{153.8}{\mega\hertz} & 63607 & \SI{0.413}{\milli\second} & 13928 Slices & 2421 ops & 0.174 ops/Slice \\
    \midrule
    ~\cite{ASSCC2019} & ASIC 65nm FDSOI & \SI{250}{\mega\hertz} & 8487 & \SI{56.2}{\micro\second}@\SI{1.1}{\volt} & \SI{12.8}{\milli\meter^2} & 17.8 kops & 1.39 kops/\unit{\milli\meter^2} \\
    Ours (1-core) & ASIC 40nm LP & \SI{769}{\mega\hertz} & 63607 & \SI{82.7}{\micro\second}@\SI{1.1}{\volt} & \SI{1.77}{\milli\meter^2} & 12.1 kops & 6.83 kops/\unit{\milli\meter^2} \\
    Ours (8-core) & ASIC 40nm LP & \SI{769}{\mega\hertz} & 63607 & \SI{82.7}{\micro\second}@\SI{1.1}{\volt} & \SI{8.00}{\milli\meter^2} & 96.7 kops & 12.09 kops/\unit{\milli\meter^2} \\
    Ours (8-core)\footnotemark[1] & ASIC 65nm (equiv.) & \SI{423}{\mega\hertz} & 63607 & \SI{150.2}{\micro\second}@\SI{1.1}{\volt} & \SI{12.0}{\milli\meter^2} & 53.3 kops & 4.44 kops/\unit{\milli\meter^2} \\
    \bottomrule
    \end{tabular}
    \footnotesize{\footnotemark[1] Row has been normalized to be equivalent to 65nm technology from 40nm LP technology~\cite{STILLMAKER201774}.}
    \end{minipage}
\end{table*}

Operator variants combination (examples in~\autoref{tab:dse-opv}) defines the mapping rule of higher-level operators into lower-level operators. As analyzed in~\hyperref[para:motivation-dse]{Section 2.2}, its influence should be considered jointly with the hardware model. In \texttt{Finesse}, the ALU configuration determines the computational capacity for executing instructions in parallel. Additionally, instruction issue width and instruction scheduling are critical components shaping the architecture's throughput and latency characteristics.

Our framework accommodates a variety of design directives, offering flexibility in the optimization process. The framework can adjust accordingly, whether the focus is on a single objective, such as maximizing throughput or minimizing area, or balancing multiple objectives in a trade-off. This capability ensures that the design process remains versatile, allowing users to define and prioritize their own performance metrics without being restricted to predefined directives. As a result, the exploration process can target performance, efficiency, device resources, or a combination of factors.

The exploration process in our framework is driven by a co-design feedback loop that iteratively refines the hardware-software configuration. This process gathers cycle info from the simulator and hardware metrics provided by EDA tools, enabling continuous optimization of both architectural and algorithmic decisions. \texttt{Finesse} incorporates basic exploration strategies, using exhaustive search for operator variants combinations. \texttt{Finesse} sets the foundation for efficiently navigating the design space, adapting to specific platform-specific constraints and optimization objectives.

\section{Evaluation}

\paragraph{\textbf{Framework Implementation}}

We implemented the \texttt{Finesse} framework in synergy of multiple ecosystems. The compiler and simulator are written in Python, supporting flexible configuration through YAML configuration files and modular invocation with command-line parameters. The parameterized hardware is implemented by SystemVerilog, with its settings automatically read from headers generated by the compilation stack. We also developed a basic operator kit containing elliptic curve operators in both Jacobian and projective coordinates, together with finite field operators from $\mathbb{F}_p$ to $\mathbb{F}_{p^{24}}$ along the finite division lattice of $24$. On top of that we implemented 7 curves in 3 curve families as listed in~\autoref{tab:curve-eg}, and validated correct functionality for all of the resulting accelerators.

\paragraph{\textbf{Hardware Validation Setup}}
The experiments were conducted on both ASIC and FPGA platforms.

\begin{itemize}
    \item \textbf{ASIC}: 40nm LP process, using a \SI{1.1}{\volt} and a \SI{25}{\degreeCelsius} typical-typical (TT) library.
    \item \textbf{FPGA}: Xilinx Virtex-7 FPGA, with 108,300 slices, 3,600 DSP blocks, and 1,470 BRAM blocks. 
\end{itemize}

\paragraph{\textbf{Performance Measurement \& Scaling}}
\label{para:eval-setup-perf}
We use standard EDA toolchain to obtain clock cycles and latency, from which throughput is calculated. The results are precise and time-deterministic, demonstrating that our design is resistant to timing attacks. Experimental results for the FPGA platform are obtained from Vivado tools, including performance and utilization.
For the ASIC implementation, performance and area metrics are derived from synthesis using commercial EDA tools.
Since different ASIC technology nodes significantly impact performance, area, etc., we refer to~\cite{STILLMAKER201774} and apply equivalent scaling adjustments for these metrics between different technology nodes for ensuring a fair comparison of pairing metrics across various ASIC implementations.

\paragraph{\textbf{Key Aspects in Evaluation}}

We evaluate the methods and contributions of \texttt{Finesse} as a framework for designing pairing-based cryptography accelerator by addressing five key aspects:
\begin{enumerate}
    \item \textit{Design efficiency}: how efficient is the \texttt{Finesse} design framework?
    \item \textit{Scalability}: is \texttt{Finesse} scalable as security level rises?
    \item \textit{Compilation optimization}: how does \texttt{Finesse}'s compile strategies improve pipeline efficiency?
    \item \textit{Co-design}: to what extent can \texttt{Finesse}'s co-design mechanism tackle the complexity of design space?
    \item \textit{Agility}: how agile and practical is the \texttt{Finesse} design framework?
\end{enumerate}

\subsection{How efficient is the Finesse design framework?}

To assess the efficiency of \texttt{Finesse}, we compare our approach with two representative works:~\cite{bag2021flexipair} and~\cite{ASSCC2019}, with the details presented in~\autoref{tab:prf-total}. The work in~\cite{bag2021flexipair} utilizes the BN256 curve, while~\cite{ASSCC2019} employs the BN254 curve, with both offering an equivalent security level. For consistency, we select the BN254 curve as our test case.

Flexipair~\cite{bag2021flexipair} stands out for its lightweight, programmable nature, providing high flexibility. However, its limited abstraction constrains extensibility. The lack of abstraction leads to an inability to achieve effective performance exploration. By contrast, \texttt{Finesse} demonstrates a 34$\times$ performance improvement, utilizing 5.6$\times$ the resources relative to~\cite{bag2021flexipair} and significantly improving area (slice) efficiency to 6.2$\times$. On the ASIC platform, \cite{ASSCC2019} delivers SOTA performance due to a highly customized ALU design, but sacrifices flexibility. By contrast, \texttt{Finesse} achieves a 3$\times$ throughput increase and 3.2$\times$ area efficiency improvement in an 8-core configuration.

This comparison shows that the \texttt{Finesse} Framework excels in both performance and area/resource efficiency. Specifically, it not only delivers significant performance gains but also demonstrates superior resource utilization. In conclusion, the \texttt{Finesse} design framework is highly efficient for pairing accelerator design, offering a high performance and resource-efficient solution.

\subsection{Is Finesse scalable as security level rises?}

\begin{figure}[t]
    \centering
    \includegraphics[width=\columnwidth]{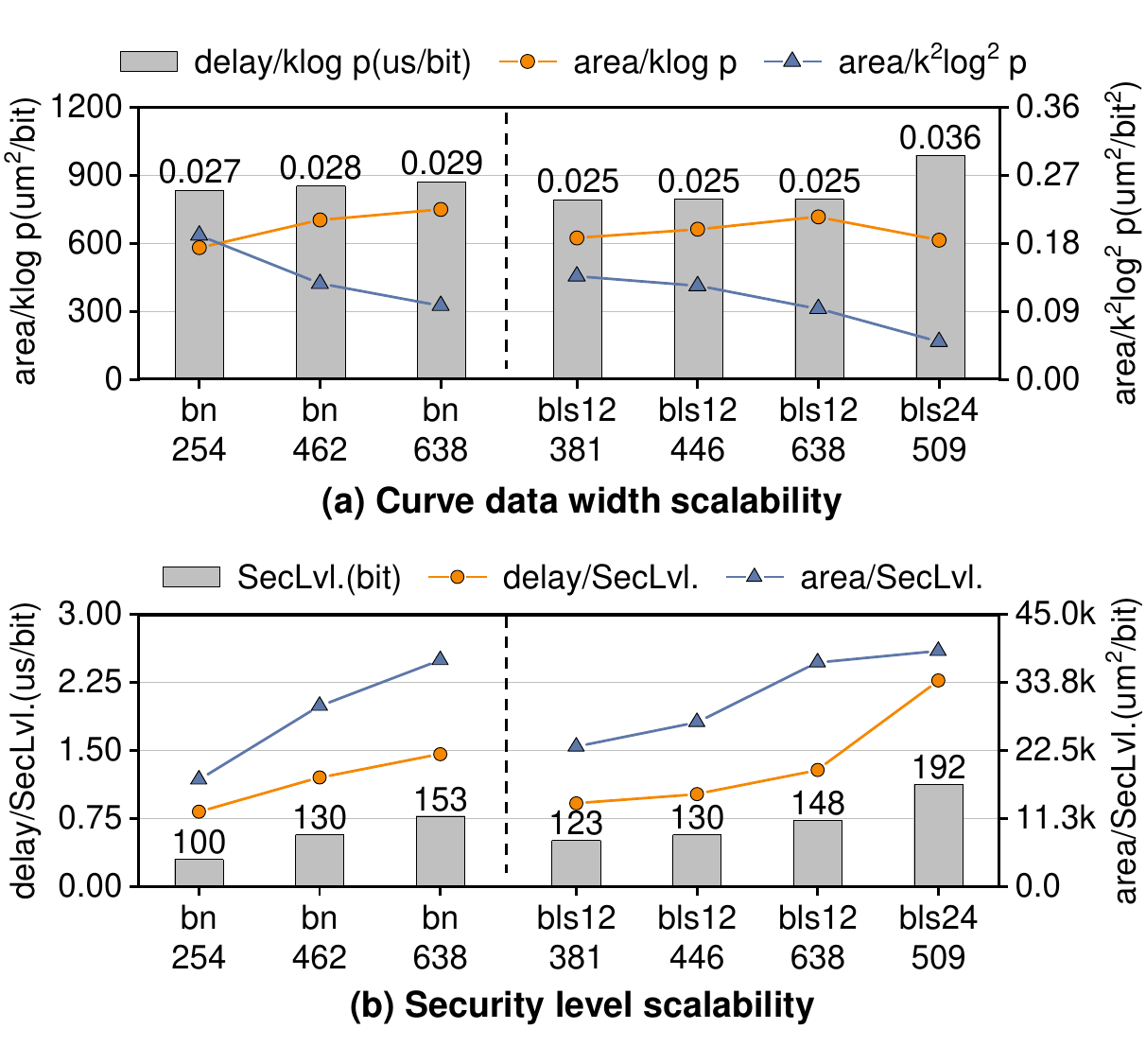}
    \caption{Scalability evaluation of \texttt{Finesse} framework.
    Respectively, $\log p$, SecLvl. and $k$ refer to the base field bit-width, security level and embedding degree. 
    }
    \label{fig:scalability}
    \Description{Fully described in the caption.}
\end{figure}

Recent progress in number field sieve (NFS) has shaken the security of pairing-based cryptography~\cite{wang2023security}. To maintain adequate security, cryptography systems must adopt larger bit-width curves and increase the embedding degree, which strains hardware resources.

\texttt{Finesse} design framework offers scalability to address these challenges posed by increasing the curve security level. \autoref{fig:scalability} illustrate the performance and scalability of the framework across different curve configurations.

In \autoref{fig:scalability}(a), we present the relationship between area, delay, and $ k \log p $ for two curve families. The pairing delay exhibits approximately linear growth as the $ k \log p $ increases. The ratios of area to $ k \log p $ and $ k^2 \log^2 p $ are plotted, indicating that despite the increase in computational complexity, the framework controls the area growth to slightly above linear, significantly below the quadratic growth rate anticipated by the complexity of finite field multiplication.

In \autoref{fig:scalability}(b), we use the method proposed by Barbulescu and Duquesne~\cite{barbulescu2019updating} to evaluate curve security under the SexTNFS~\cite{extnfs2016} attack. The evaluation results are presented in the gray bar plot, indicating an increase in security level as the $ k \log p $ expands. Meanwhile, the line chart indicates that as the security level increases, the ratio of pairing delay to security level remains relatively stable, while the area growth is kept within a reasonable range.

The results suggest that the \texttt{Finesse} framework is scalable as security levels increase and can effectively maintain a balance between performance and resource consumption.

\subsection{How does Finesse's compile strategies improve pipeline efficiency?}

\label{para:eval-compile}

\begin{figure}[t]
    \centering
    \includegraphics[width=\linewidth]{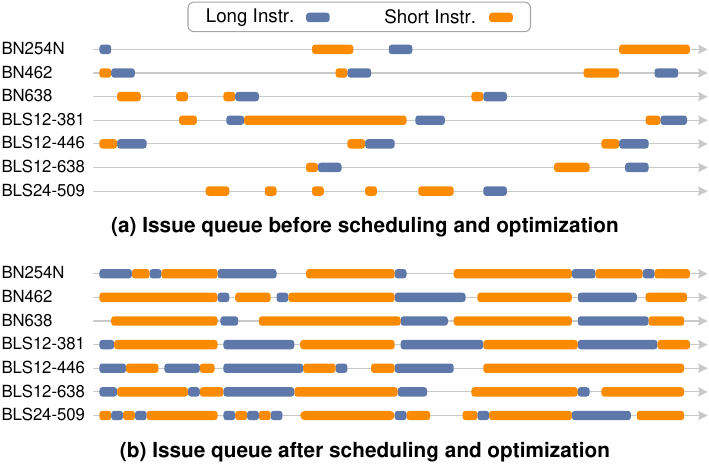}
    \caption{Scheduling and issue slot affinity optimization on issue queue. Each instruction occupies a single slot in issue queue. The snapshot is taken starting from the 10,000th cycle during the simulated execution of SSA IR. Hardware model: \texttt{Long} = 38cy, \texttt{Short} = 8cy, \texttt{2R1W}/cy. \#Curve: 7. Algo: O-Ate.}
    \Description[]{Fully described in the caption.}
    \label{fig:eval-sched-opt}
\end{figure}

\texttt{Finesse}'s compile strategy directly addresses pipeline efficiency through a combination of data flow and architecture-specific scheduling optimizations.

Finding a suitable compilation baseline for \textbf{emerging workloads} on a novel customized target accelerator is a non-trivial task. The architectural diversity among accelerators presents a significant challenge for establishing a common compilation baseline. For example, \cite{bag2021flexipair} is CISC-like with sub-$\mathbb{F}_p$ ALUs, whereas \cite{ASSCC2019} is FSM-like with $\mathbb{F}_{p^2}$ ALUs. There exists no widely accepted methodology for precisely and fairly comparing instruction count or cycle count across different accelerator architectures. Macro-level comparison of compilation effect is not possible without a common target or established equivalance relation between targets.

\begin{table}[t]
    \centering
    \begin{minipage}{\linewidth}
    \caption{Evaluation of \texttt{Finesse}'s compilation strategies. Compile time ranges from 8.0s/BN254N to 53.1s/BLS24-509. Hardware model: \texttt{Long} = 38cy, \texttt{Short} = 8cy, \texttt{2R1W}/cy. Algo: O-Ate.}
    \label{tab:ssa-opt}
    \begin{tabular}{lcc}
        \toprule
        \multirow{2}{*}{\textbf{Curve}} & \multicolumn{1}{c}{\textbf{Instr. Reduction}} & \multicolumn{1}{c}{\textbf{IPC Improvement}} \\
        & \textbf{Init. $\to$ Opt.} & \textbf{Init. $\to$ Opt. (HW 1/2)\footnotemark[1]} \\
        \midrule
        BN254N & 62.7k $\to$ 55.3k (-11.7\%) & 0.19 $\to$ 0.87 / 0.92\\
        BN462 & 115k $\to$ 101k (-12.0\%) & 0.20 $\to$ 0.88 / 0.92\\
        BN638 & 155k $\to$ 137k (-12.0\%) & 0.20 $\to$ 0.88 / 0.92\\
        BLS12-381 & 81.1k $\to$ 74.1k (-8.73\%) & 0.19 $\to$ 0.87 / 0.92\\
        BLS12-446 & 94.8k $\to$ 86.5k (-8.73\%) & 0.19 $\to$ 0.87 / 0.92\\
        BLS12-638 & 125k $\to$ 114k (-8.47\%) & 0.19 $\to$ 0.87 / 0.92\\
        BLS24-509 & 324k $\to$ 271k (-16.4\%) & 0.22 $\to$ 0.88 / 0.97\\
        \bottomrule
    \end{tabular}
    \footnotesize{\footnotemark[1] HW 1/2 refers to the hardware model without/with FIFO buffer, which is an architectural feature alleviating write-back conflicts.}
    \end{minipage}
\end{table}

\begin{figure}[t]
    \centering
    \includegraphics[width=\columnwidth]{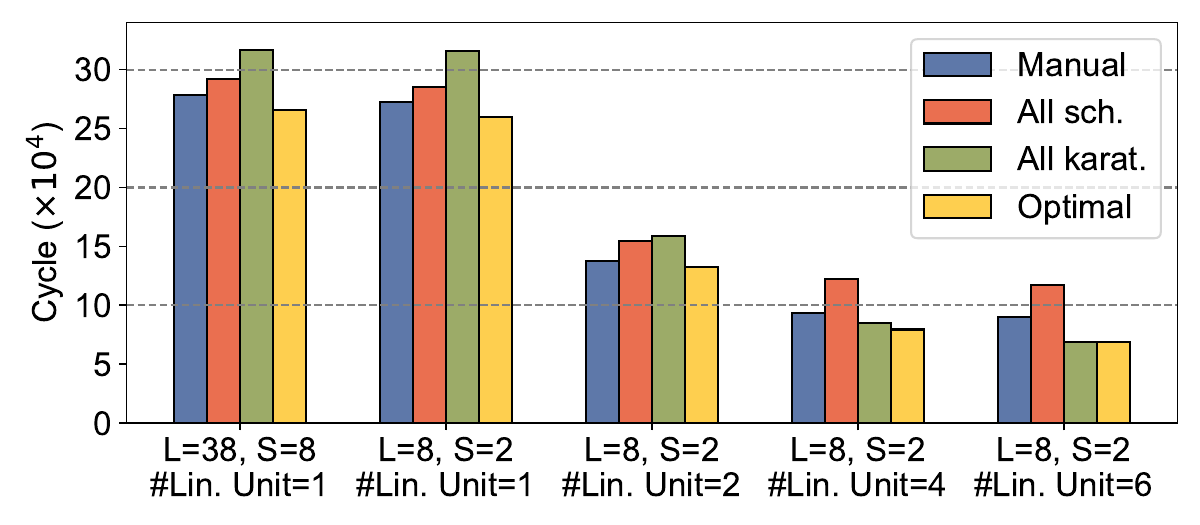}
    \caption{\texttt{Finesse}'s design space search on operator variant combinations and representative pipeline configurations. ``Manual'' refers to combinations of variants we selected manually; ``All sch.'' and ``All karat.'' refer to the combinations entirely using the Schoolbook and Karatsuba variants, respectively; ``Optimal'' refers to the best combinations obtained from exploration. ``L'' and ``S'' denote the cycles for \texttt{long} and \texttt{short} instructions, and ``\#Lin. Unit'' indicates the number of linear units, which also equals the number of register banks. Hardware model: \texttt{Long} = Lcy, \texttt{Short} = Scy, \texttt{2R1W}/bank/cy. Curve: BLS24-509, Algo: O-Ate.}
    \Description{Fully described in the caption.}
    \label{fig:eval-dse-operator}
\end{figure}

\textbf{Our baseline} (referred to as ``Init.'' in~\autoref{tab:ssa-opt}, ``before'' in~\autoref{fig:eval-sched-opt}, ``Manual'' in~\autoref{fig:eval-dse-operator}) pairing implementation is built directly from cryptographic literature, i.e. exactly as reported, without alterations that might introduce a favorable bias towards our compiler.

Through data flow optimizations, \texttt{Finesse} automates transformations that were previously handled manually in research, such as $\mathbb{F}_{p^{k}}$ \textit{dense} $\times$ \textit{sparse} multiplication. \autoref{tab:ssa-opt} quantifies their impact in terms of instruction reduction across multiple curves. Rather than claiming credit for the optimization itself, \Finesse~contributes to agility by performs these optimizations transparently, freeing users from manually handling sparsity.

In terms of scheduling optimization, \texttt{Finesse} employs novel scheduling strategies, which include standard code motion based on instruction latencies, together with instruction issue slot affinity optimization. \autoref{fig:eval-sched-opt} gives a clear visualization in the form of a waterfall chart, illustrating the improved pipeline utilization, showing how pipeline bubbles are minimized. \autoref{tab:ssa-opt} contains detailed IPC statistics.

These results have demonstrated that \texttt{Finesse}'s optimization strategies significantly improve pipeline efficiency, resulting in notable performance gains in pairing accelerators through reducing execution time and enhancing resource utilization. Collectively, these optimizations provide a solid building block for \texttt{Finesse}'s co-design mechanism.

\subsection{To what extent can Finesse's co-design mechanism tackle the complexity of design space?}
\label{para:eval-dse}

\texttt{Finesse}'s co-design mechanism is an initial step towards full-fledged design space exploration. Our work serves as a stepping stone from 0 to 1, paving the way for a more comprehensive approach to optimizing performance across a vast design space.

\texttt{Finesse} provides a fully functional implementation capable of performing exhaustive design space exploration. \autoref{fig:eval-dse-operator} presents results from \texttt{Finesse}'s analysis of operator variant combinations and representative pipeline configurations. Unlike the conventional approach of applying Karatsuba optimization at all levels, we explored a variety of operator variants for the BLS24-509 curve, combined with hardware models ranging from a basic single-issue pipeline to a multi-issue pipeline featuring up to 6 linear modular arithmetic units. In addition to typical baseline approaches, we included a manually selected variant combination, guided by heuristics optimized for a single-issue pipeline. With limited parallelism in linear operations this manually tweaked version outperforms typical approach and is near optimal, but with more linear units, all-Karatsuba is still a viable choice.

\begin{figure}[t]
    \centering
    \includegraphics[width=\columnwidth]{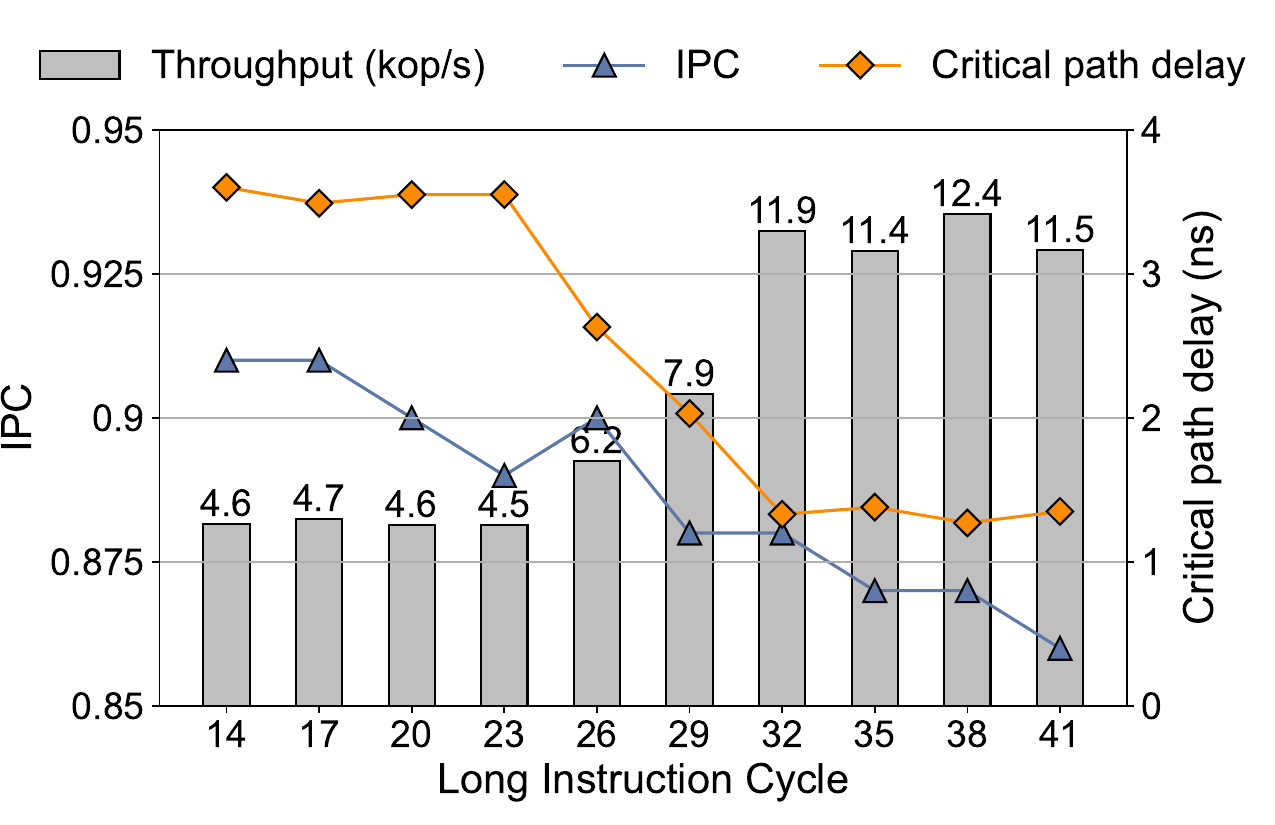}
    \caption{\texttt{Finesse}'s co-design mechanism with respect to choices in ALU family. Hardware model: \texttt{Long} = $x$cy, \texttt{Short} = 8cy, \texttt{2R1W}/cy. Curve: BN254N, Algo: O-Ate.}
    \Description{Fully described in the caption.}
    \label{fig:eval-dse-alu}
\end{figure}

\texttt{Finesse} can also perform co-design with feedbacks from EDA toolchain. \autoref{fig:eval-dse-alu} reflects the impact of choices regarding the ALU family (ALU family refers to the fully pipelined mmul units with different pipeline depths, which are equivalent to the \texttt{Long} instruction cycles described in the paper), with ALU design variations being the primary parameter for optimization. The ALU critical path information is gathered from synthesis results of ASIC toolchain over target technology node, while compile-time estimates of IPC and throughput metrics are obtained from the simulator. These metrics are derived from \texttt{Finesse}'s optimization pass, which utilizes hardware abstractions passed in by the co-design mechanism as essential directives in IR scheduling.

The results in \autoref{fig:eval-dse-alu} indicate a drop in IPC with deeper ALU pipelines, due to the limited inherent parallelizability of the O-Ate pairing algorithm. Additionally, as determined by the constraints of the target technology node, critical paths cease to decrease with deeper pipelines. \texttt{Finesse} effectively analyzes this real-world non-linear relationship, identifying the optimal pipeline depth of 38 cycles on single issue architectures for our experiment setup.

\subsection{How agile and practical is the Finesse design framework?}

The \texttt{Finesse} design framework offers both agility and practicality. By automating a large part of the design cycle and supporting a wide range of configurations, \texttt{Finesse} allows users to swiftly adapt to diverse application needs. \texttt{Finesse}'s effectiveness is best illustrated through various use case scenarios and security considerations.

\paragraph{\textbf{For Pairing Researchers}}
\texttt{Finesse}'s basic operator kit enables rapid porting of pairing algorithms, allowing users to quickly experiment with new approaches to pairing constructions or novel families of curves, receiving architectural feedback in just minutes. This capability significantly shortens the design cycle, enabling the swift creation of well-optimized hardware accelerators tailored to innovative ideas in pairing-based cryptography.

\begin{figure}[t]
    \centering
    \includegraphics[width=\columnwidth]{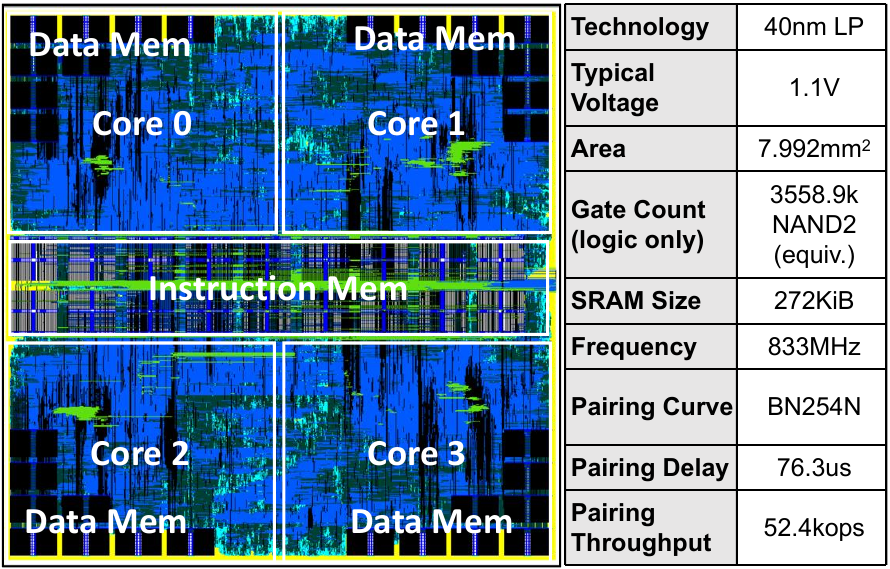}
    \caption{Quad-core \texttt{Finesse} chip layout. Note that timing result is slightly better than synthesis results.}
    \Description{Fully described in the caption.}
    \label{fig:eval-layout}
\end{figure}

\paragraph{\textbf{For Hardware Accelerator Designers}}
\texttt{Finesse} provides ready-for-use implementations of hardware accelerators in SystemVerilog out of the box, supports generating information aiding RTL-level behavioral simulation. Designers are free to extend and experiment with more advanced ALUs and storage blocks using \texttt{Finesse}, given that it can be incorporated into the hardware abstraction system of \texttt{Finesse}. \texttt{Finesse}'s default implementation utilizes standard technology cells, supported by a broad range of EDA toolchains and platforms. \texttt{Finesse}-generated designs are compatible with standard EDA flows, including synthesis, layout, and validation. \autoref{fig:eval-layout} gives an experimental ASIC layout of quad-core accelerator designed by \texttt{Finesse}, showcasing the practicality and effectiveness of the final product. The framework allows for rapid adjustments to the hardware architecture, further enhancing design agility.

\paragraph{\textbf{Security Considerations}}

\label{para:eval-security}

\Finesse~provides support at the level of IP cores, meaning it is typically integrated with other user-defined peripheral IPs to form a complete IC die or be deployed on an FPGA board. The physical security of the chip, whether ASIC or FPGA, largely depends on its specific deployment setup. A rigorous and comprehensive evaluation of this aspect is beyond the scope of this paper, which focuses on architecture and design automation.

Nevertheless, at the current level of support, we can qualitatively assess its resilience against classic attacks, such as basic side-channel and fault-injection attacks. By design, \Finesse~is inherently resistant to timing attacks, as pairing computations are designed to complete in a fixed number of cycles. Additionally, its instr/data fetch patterns are independent of sensitive inputs, providing a solid basis against attacks that exploit data access patterns. Regarding fault-injection attacks, a bit-flip in the program counter could potentially leak low-rank information about sensitive data. However, this risk can be mitigated by introducing redundancy and/or error correction mechanisms in key memory modules within \Finesse.

In conclusion, these use case scenarios and security considerations collectively demonstrate how the \texttt{Finesse} design framework combines agility with practicality, streamlining the design process from algorithm development to final layout and ensuring responsiveness to the evolving demands of pairing-based cryptography.

\section{Future Works}

\paragraph{\textbf{Next steps for design space exploration}}
\label{para:dse-future}

Our abstraction system supports different memory organization schemes (such as bank configurations) under VLIW. Once hardware support for VLIW is implemented (which is essentially an engineering task), its performance data can be incorporated into the DSE cycle. The choice of memory bank partitioning schemes impacts the memory area at the hardware level, while the number of memory banks affects bank conflict rates during software compilation. This addition introduces a new dimension to the design space, which drives us to pursue more efficient searching strategies.
In fact, our parameter space exhibits well-defined adjacency (e.g. adjacent nodes on hypercube), which is suitable for advanced strategies (e.g. simulated annealing).

\paragraph{\textbf{Constructing a GEM5 model to enhance framework efficiency}}
\label{para:future-gem5}
In the future work, we intend to develop a equivalent model utilizing GEM5 to improve the efficiency of the \Finesse~framework. As an open-source system emulator, GEM5 provides a comprehensive simulation of hardware behavior and performance evaluation, enabling us to accurately evaluate hardware performance and power consumption during the design phase. Meanwhile, our open sourced \Finesse~framework will be upgraded to be compatible with the equivalent GEM5 model.

\paragraph{\textbf{Supporting wider range of cryptographic algorithms}}
\label{para:future-general}
The hierarchical abstraction of the \Finesse~enables us to agilely extend the framework to support broader range of classical cryptographic algorithms. For instance, to implement a block cipher algorithm such as AES, it is sufficient to integrate fixed-length octet-stream data operation instructions and type conversion instructions between the $\mathbb{F}_p$ and octet-stream into the ISA.

\section{Conclusion}
 
\texttt{Finesse} is an agile design framework for pairing-based cryptography accelerators, providing a novel abstraction that supports automated exploration across multiple layers, spanning algorithms, operators, programs, and hardware, as well as facilitating cross-layer co-design. Through effective co-design, \texttt{Finesse}'s accelerator significantly outperforms state-of-the-art solutions in both performance and area efficiency. The agility of \texttt{Finesse} is reflected in its flexible configuration, automated support for both process and optimization, streamlined multi-disciplinary collaboration, and efficient facilitation of the design process for pairing accelerators. By allowing researchers from the algorithm, compiler, and hardware domains to focus on their specific expertise without needing to understand the complete system, \texttt{Finesse} significantly reduces learning and development barriers, making it an effective approach for cryptographic accelerator design.

\bibliographystyle{ACM-Reference-Format}
\bibliography{references}

\end{document}